\documentclass{ws-procs9x6}
\include{psfig.tex}
\begin{document}
\newcommand{\asfigure}[3]{\psfig{#1}}
\def\cH{{\it H}}
\def\cP{{\it P}}
\def\cD{{\it D}}
\def\cG{{\it G}}
\def\cV{{\it V}}
\def\cF{{\it F}}
\def\cU{{\it U}}
\def\cS{{\it S}}
\def\cO{{\it O}}
\def\cE{{\it E}}
\def\bfA{{\bf A}}
\def\bfG{{\bf G}}
\def\bfn{{\bf n}}
\def\bfr{{\bf r}}
\def\bfV{{\bf V}}
\def\bft{{\bf t}}
\def\bfM{{\bf M}}
\def\bfP{{\bf P}}
\def\bra#1{\langle #1 |}
\def\ket#1{| #1 \rangle}
\newcommand{\p}{\partial}
\def\coltwovector#1#2{\left({#1\atop#2}\right)}
\def\upp{\coltwovector10}
\def\downn{\coltwovector01}
\def\Ord#1{{\it O}\left( #1\right)}
\def\bmp{\mbox{\boldmath $p$}}
\def\rhobar{\bar{\rho}}
\renewcommand{\Re}{{\rm Re}}
\renewcommand{\Im}{{\rm Im}}
\renewcommand{\theequation}{\thesection.\arabic{equation}}
\title{The Optical Approach to Casimir Effects}
\author{A.~Scardicchio\footnote{\uppercase{J}oint work with
\uppercase{R.~L.~J}affe}}
\address{Center for Theoretical Physics, \\ Laboratory for
   Nuclear Science and Department of Physics \\ Massachusetts
Institute
  of Technology \\ Cambridge, MA 02139, USA}
\maketitle

\abstracts{We propose a new approach to the Casimir effect based
on
    classical ray optics.  We define and compute the contribution of
    classical optical paths to the Casimir force between rigid bodies.
    Our
    approach improves upon the proximity force approximation.  It can be
    generalized easily to arbitrary geometries, different boundary conditions,
    to the computation of Casimir energy densities and to many other
    situations. \\ \\
    MIT-CTP-3522}

\section{Introduction}
\setcounter{equation}{0} \setcounter{figure}{0}

The last 10 years have witnessed quite a revolution in the
experimental techniques used to prove Casimir effect.\cite{expt1}
Casimir's original prediction for the force between grounded
conducting plates due to modifications of the zero point energy of
the electromagnetic field has already been verified to an accuracy
of a few percent. Progress has been slower on the theoretical
side. Beyond Casimir's original study of parallel
plates,\cite{Casimir} we are only aware of useful calculations for
a corrugated plate\cite{Kardar} and for a sphere and a
plate.\cite{Gies03} Simple and experimentally interesting
geometries like two spheres, a finite inclined plane opposite an
infinite plane, and a pencil point and a plane, remain elusive.
The Proximity Force Approximation\cite{Derjagin} (PFA) was shown
by Gies {\it et al.\/}\cite{Gies03} to deviate significantly from
their precise numerical result for the sphere and plane. Thus at
present neither exact results nor reliable approximations are
available for generic geometries. It was in this context that we
recently proposed a new approach to Casimir effects based on
classical optics.\cite{pap1} The basic idea is extremely simple:
first the Casimir energy is recast as a trace of the Green's
function; then the Green's function is approximated by the sum
over contributions from optical paths labelled by the number of
(specular) reflections from the conducting surfaces. The integral
over the wave numbers of zero point fluctuations can be performed
analytically, leaving a formula which depends only on the
properties of the paths between the surfaces. This approach will
give an approximation (though a surprisingly accurate one) which
is valid when the natural scales of diffraction are large compared
to the scales that measure the strength of the Casimir force.  In
practice this will typically be measured by the ratio of the
separation between the conductors, $a$, to their curvature, $R$.
It generalizes naturally to the study of Casimir thermodynamics,
energy and pressure, to various b.c., to fermions, and to compact
and/or curved manifolds.

In the optical approximation the cutoff dependent terms of the
Casimir energy can easily be isolated and shown to be independent
of the separation between conductors. They therefore do not
contribute to forces and can be dropped.

\section{Derivation}
\label{sec:deriv}
\setcounter{equation}{0}

Most studies of Casimir energies do not consider approximations.
Instead they focus on ways to regulate and compute the sum over
modes, $\sum \frac{1}{2}\hbar\omega$.\cite{MT}  These methods have
proved very difficult to apply to geometries other than parallel
plates.  The main reason for this \emph{impasse} lies in the
requirement of an analytic knowledge of the spectrum of the
Laplace operator for the given geometry. However, the knowledge of
this spectrum for a family of boundaries (like a sphere facing a
plane) would have the most non-trivial implications for the same
family of quantum billiards and hence of classical billiards. 30
years of work on the ergodicity of classical billiards and their
quantum counterparts suggest this task is hopeless.\cite{Gutz}

A numerical knowledge of the spectrum does not represent a
reliable solution to the problem either. The force, indeed, is
given by the small oscillatory ripple in the density of state
numerically shadowed by the `bulk' contributions which give rise
to distance-independent divergencies.

So we focused our attention on ways to get approximate solutions
of the Laplace-Dirichlet problem which are apt to capture the
oscillatory contributions in the density of states, providing
physical insights and accurate numerical estimates.

\subsection{The Optical Approximation for the propagator}
There is a strong connection between the Casimir energy for a
field $\phi$ obeying Dirichlet boundary conditions (b.c.) on the
boundary of the domain $\cD$ and the propagator $G(x',x,k)$ of the
Helmholtz equation on the same domain with the same b.c.\/ . The
knowledge of the latter allows one to calculate the density of
states $\rho(k)$ and from this we can obtain the Casimir energy by
quadratures. Indeed, from the well-known definition of the Casimir
energy in terms of a space and wave-number dependent density of
states,\cite{densityofstates} $\rho(x,k)$,
\begin{eqnarray}
    \label{eq:casimiren}
    \cE_{\cD}[\phi]=\int_0^\infty dk \int_{\cD} d^{N} x
    \frac{1}{2}\hbar\ \omega(k)\rho(x,k),
\end{eqnarray}
where $\omega(k)=c\sqrt{k^{2}+\mu^{2}}$, and the density of states
$\rho(x,k)$ is related to the propagator $G(x',x,k)$ by
\begin{equation}
    \label{eq:rhoG}
    \rho(x,k)=\frac{2k}{\pi}\ \Im \ G(x,x,k).
\end{equation}
where the usual density of states is $\rho(k)=\int d^N
x\rho(x,k)$.

We must choose $G$ to be analytic in the upper-half $k^{2}$-plane;
in the time domain (see later) this means we are taking the
retarded propagator.

The Casimir energy depends on the b.c.\/ obeyed by the field
$\phi$ and on the arrangement of the boundaries,
$\cS\equiv\partial \cD$ (not necessarily finite), of the domain
$\cD$.  From the outset we recognize that $\cE$ must be regulated,
and will in general be cutoff dependent.  We will not denote the
cutoff dependence explicitly except when necessary. $\rho$ and $G$
are the familiar density of states and propagator associated with
the problem
\begin{equation}
\label{eq:Cauchyprob}
    (\Delta+k^2)\psi(x)=0\mbox{\quad for\quad}x\in\cD,\quad
    \psi(x)=0\mbox{\quad for\quad}x\in\cS.
\end{equation}
We can regard this problem as the study of a quantum mechanical
free particle with $\hbar=1$, mass $m=1/2$, and energy $E=k^{2}$,
living in the domain $\cD$ with Dirichlet b.c.\/ on $\partial
\cD$. Dirichlet b.c.\/ are an idealization for the interactions
which prevent the quantum particle from penetrating beyond the
surfaces $\cS$.  This is adequate for low energies but fails for
the divergent, {\it i.e.\/} cutoff dependent, contributions to the
Casimir energy.\cite{Graham:2003ib} However, the divergences can
be simply disposed of in the optical approach, and the physically
measurable contributions to Casimir effect are dominated by $k\sim
1/a$, where $a$, a typical plate separation, satisfies $1/a\ll
\Lambda$ whit $\Lambda$ being the momentum cutoff characterizing
the material.  So the boundary conditions idealization is quite
adequate for our purposes. Following this quantum mechanics
analogy we introduce a fictitious time, $t$, and consider the
functional integral representation of the
propagator.\cite{Schulman}  The space-time propagator $G(x',x,t)$
obeys the free Schr\"odinger equation in $\cD$ bounded by $\cS$.
It can be written as a functional integral over paths from $x'$ to
$x$ with action $S(x',x,t)=\frac{1}{4}\int_0^t dt\dot x^{2}$. The
optical approximation is obtained by taking the stationary phase
approximation of the propagator $G$ \emph{in the fictitious time
domain}
\begin{equation}
\label{eq:tsemicl}
    G_{{\rm opt}}(x',x,t)=\sum_{r}K_r(x',x,t)e^{i
    S_r(x'x,t)}.
\end{equation}
The classical action is
\begin{equation}
    S_r(x',x,t)=\frac{\ell_r(x',x)^2}{4t}
\end{equation}
and $K_r$ is the van Vleck determinant
\begin{equation}
    K_r(x',x,t)\propto \det\left(\frac{\partial^2 \ell_r^2}{\partial
    x'_i\partial x_j}\right)^{1/2}.
\end{equation}
With some manipulations\cite{pap1} we can turn this determinant
into
\begin{equation}
    K_r(x',x,t)=\frac{(-1)^{r}}{(4\pi i
    t)^{N/2}}\left(\ell_r^{N-1}\frac{d\Omega_x}{dA_x'}\right)^{1/2},
\end{equation}
where $N$ is the number of spatial dimensions. This approximation
is exact to the extent one can assume the classical action of the
path $S_r$ to be quadratic in $x',x$. This is the case for flat
and infinite plates.  Thus the non-quadratic part of the classical
action comes from the curvature or the finite extent of the
boundaries, which we parameterize generically by $R$, $\partial^3
S/\partial x^3\sim 1/Rt$.

In $k$-space the corrections hence will be $\Ord{1/kR}$, and the
important values of $k$ for the Casimir energy are of order $1/a$,
where $a$ is a measure of the separation between the surfaces.
Thus the figure of merit for the optical approximation is $
{a/R}$.  Certainly some of the curvature effects are captured by
the van Vleck determinant but at the moment there is no good way
to estimate the order in $a/R$ of the corrections to the optical
approximation (possibly fractional, plus exponentially small
terms). This is topic for further investigation.

Putting all together we find the space-time form of the optical
propagator to be
\begin{equation}
    \label{eq:tdomain}
    G_{\rm opt}(x',x,t)=\sum_{ r}\frac{(-1)^{r}}{(4\pi i
    t)^{N/2}}\left(\ell_r^{N-1}\frac{d\Omega_x}
    {dA_x'}\right)^{1/2}e^{i\ell_r^2/4t}.
\end{equation}
When dealing with infinite, parallel, flat plates this
approximation becomes exact.  For a single infinite plate, for
example, the length-squared of the only two paths going from $x$
to $x'$ are $\ell_{{\rm direct}}^2=||x'-x||^2,\quad
    \ell_{{\rm 1 reflection}}^2=||x'-\tilde{x}||^2$,
where $\tilde x$ is the image of $x$.  Both are quadratic
functions of the points $x,x'$ and the optical approximation is
indeed exact. $G_{\rm opt}(x',x,k)$ is obtained by Fourier
transformation and can be expressed in terms of Hankel functions,
giving us the final form for our approximation
\begin{eqnarray}
    G_ {\rm opt}(x',x,k)&=&\sum_r\frac{(-1)^{r}i\pi}{(4\pi
    )^{N/2}}\left(\ell_r^{N-1}\Delta_r
    \right)^{1/2}\left(\frac{\ell_r}{2k}\right)^{1-N/2}H^{(1)}_{\frac{N}{2}-1}
    \left(k\ell_r\right),\nonumber \\
    \label{eq:unifsemicl}
    &\equiv&\sum_r G_r(x',x,k),
\end{eqnarray}
where $\Delta_r$ is the enlargement factor (see\cite{pap1} for
details)
\begin{equation}
    \label{eq:deltar}
    \Delta_r(x',x)=\frac{d\Omega_{x}}{dA_{x'}}
\end{equation}
and we have suppressed the arguments $x$ and $x'$ on $\ell_{r}$
and $\Delta_{r}$ in (\ref{eq:unifsemicl}).  This can be thought of
as a particular case of a more general result.\cite{Berry72}

\subsection{The Optical Casimir energy}
The substitution of (\ref{eq:unifsemicl}) into (\ref{eq:rhoG}) and
then in (\ref{eq:casimiren}) gives rise to a series expansion of
the Casimir energy associated with classical closed (but not
necessarily periodic) paths
\begin{equation}
\label{eq:optser}
\cE_{\rm opt}=\sum_{{\rm paths\ } r} \cE_r,
\end{equation}
where each term of this series will be in the form of
\begin{equation}
\label{eq:rcontrib}
\cE_r=\frac{1}{2}\hbar\ \Im\int_0^\infty dk \omega(k)
\frac{2k}{\pi}\int_{\cD_r}d^Nx\ \ G_r(x,x,k).
\end{equation}
Here the integration has been restricted to the domain
$\cD_r\subset \cD$ where the given classical path $r$ exists. If
the length of the path is bounded from below (this is the case for
more than one reflection) the $x$ and $k$ integral can be switched
safely. The $k$-integral can be performed exactly for any $N$ and
$\mu$, but it is particularly simple for the massless case,
$\omega(k)=ck$,
\begin{equation}
    \cE_r=\hbar
    c\frac{(-1)^{r+1}}{2\pi^{N/2+1/2}}\Gamma\left(
    \frac{N+1}{2}\right)
    \int_{\cD_r}d^Nx\frac{\Delta_r^{1/2}}{\ell_r^{(N+3)/2}}.
    \label{ecasN}
\end{equation}
This is the central result of our work and associates a Casimir
energy contribution to each optical path $r>1$. The series
(\ref{eq:optser}) has a very fast convergence, usually $98\%$ of
the contribution is contained in the first 4 terms, as we will see
in the examples below.

\subsection{Divergencies}
The paths $r=1B$ that bounce only once on a given body $B$ must be
treated with particular care because their contribution is
divergent. The $x$ and $k$ integrals cannot be inverted without
regulating the divergencies. To do so we insert a simple
exponential cutoff in $k$. For a massless field the
$k$-integration in $\cE_{1B}$ can be performed\footnote{For
simplicity we specialize to $N=3$ although the analysis is
completely general.} giving
\begin{equation}
\cE_{1B}=-\frac{\hbar c}{4\pi^{2}} \int_{\cD_{1B}}d^3x\
\Delta_{1B}^{1/2}
\frac{2\ell_{1B}\Lambda^4(3-(\ell_{1B}\Lambda)^2)}
{(1+(\ell_{1B}\Lambda)^2)^3}.
\label{1refldiv}
\end{equation}
Notice that for $\ell_{1B}\Lambda\gg 1$ we reobtain the standard
result, eq.~(\ref{ecasN}). Hence it is convenient to rewrite this
integral as
\begin{equation}
\cE_{1B}=-\frac{\hbar c}{4\pi^{2}}\left(
\int_{\cD_{1B}\bigcup\overline{\cD}_{1B}}d^3x-\int_{\overline{\cD}_{1B}}d^3x\right)
\Delta_{1B}^{1/2}
\frac{2\ell_{1B}\Lambda^4(3-(\ell_{1B}\Lambda)^2)}
{(1+(\ell_{1B}\Lambda)^2)^3},
\label{1refldiv2}
\end{equation}
where $\overline{\cD}_{1B}$ is the domain where the path
\emph{does not} exist. The first integral does not depend on the
position of the other bodies and in the second we can take the
limit $\Lambda\to\infty$ safely because $\ell_{1B}(x)>2a>0$, where
$a$ is the minimum distance between the bodies. Hence we will
write $\cE_{1B}=\cE_{1B, {\rm div}}+\cE_{1B, {\rm fin}}$, the
finite part being the integral over $\overline{\cD}_{1B}$ (notice
the extra minus sign).

The first term contains all the divergencies arising for
$\ell_{1B}\to 0$. Notice that since when
$\ell_{1B}\Lambda=\sqrt{3}$ the \emph{sign} of the integrand
changes, the divergence is negative rather than positive, as one
could have argued from (\ref{ecasN}).\cite{Graham:2002xq} Of
course the bulk contribution to the vacuum fluctuation energy
comes from the zero-reflection term, which is positive.

Using the expression of $\Delta_{1B}$ for a general surface with
principal radii of curvature $R_{a,b}$ we find
\begin{equation}
    \cE_{1B, {\rm div}}\sim-\frac{S}{8\pi}\hbar c\Lambda^3-\Lambda^2\frac{1}{32\pi^2}\hbar c\int
    dS\left(\frac{1}{R_a}+\frac{1}{R_b}\right)+\Ord{\ln
    \Lambda}.
    \label{asymptdiv}
\end{equation}
These terms do not contribute to the forces between rigid objects.
The form of eq.~(\ref{asymptdiv}) invites comparison with the work
of Balian and Bloch.\cite{BalianBloch} One finds agreement in the
surface terms but not in the curvature terms.

\section{Parallel Plates}
\label{sec:parplat}
Parallel plates provide a simple, pedagogical example which has
many features --- fast convergence, trivial isolation of
divergences, dominance of the even reflections --- that occur in
all the geometries we analyzed.  We assume for simplicity that the
two plates have the same area $S$.  The relevant paths are shown
in Fig.~1 where the points $x$ and $x'$, which should be equal,
are separated for ease of viewing.  For the even paths
$\ell_{2n}(z)=2na$, $n=1,2,\ldots$, independent of $z$ (here $z$
is the distance from the lower surface).  For the odd paths
$\ell_{2n-1,\alpha}(z)= 2(n-1)a+2\zeta$, where $\zeta=z,a-z$
respectively if $\alpha=\mathrm{down,up}$ and $n=1,2,\ldots$ . For
planar boundaries the enlargement factor is given by $\Delta_{\bf
n}= 1/\ell^2_{\bf n}$.
\begin{figure}[ht]
\begin{center}
\epsfxsize=3.9in\epsfbox{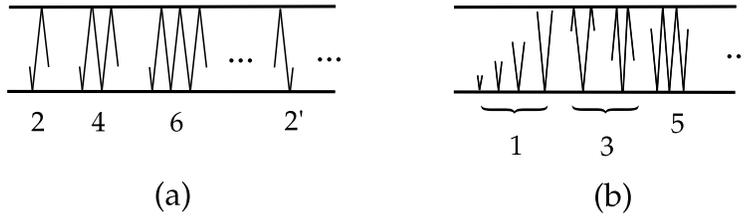} \caption{a) Even and b) odd
optical paths for parallel plates.  Initial and final points have
been separated for visibility.} \label{paths}
\end{center}
\end{figure}

The sum over even reflections,
\begin{equation}
{\cE}_{\rm even}=-\frac{\hbar c}{\pi^{2}}\sum_{n=1}^{\infty}\int
dS\int_{0}^{a}dz \frac{1}{(2na)^{4}}=-\frac{\pi^{2}\hbar c}{1440
a^{3}}S
\label{evencontr}
\end{equation}
is trivial because it is independent of $z$.  The result is the
usual Dirichlet Casimir energy.\cite{expt1}  The sum over odd
reflections, after being regulated by point splitting, before
removing the infinite part, gives
\begin{equation}
{\cE}_{\rm{odd}}=\frac{\hbar c}{2 \pi^2}\int dS\sum_{n=0}^\infty
\int_{0}^{a}dz\frac{1}{\left(\epsilon^2+(2z+2na)^2\right)^2}
=\frac{\hbar c}{16\pi^2}\frac{2\pi S}{\epsilon^3}.
\end{equation}
The divergence as $\epsilon\to 0$ is precisely what is expected on
the basis of the general analysis of the density of states in
domains with boundaries.\cite{BalianBloch,Deutsch79} Moreover,
since it is independent of $a$, it does not give rise to a force.
From Fig.~1 b) it is evident why this must happen. The total sum
of the off reflection contributions is just the integral of the
one reflection path extended up to $\infty$ and hence does not
depend on $a$.

The fact that the odd reflections sum up to a divergent constant
is universal for geometries with planar boundaries, and to a good
approximation is also valid for curved boundaries. Note also that
the sum over $n$ in eq. (\ref{evencontr}), converges rapidly:
$92\%$ of the effect comes from the first term (the two reflection
path) and $>98\%$ comes from the two and four reflection paths.
This rapid convergence persists for all the geometries we have
analyzed due to the rapid increase in the length of the paths.

\section{Sphere and Plane}
\label{sec:sphpl}

We calculated the Casimir energy for the sphere and the plane up
to four reflections. $E_1$ and $E_3$ can be found analytically
while $E_2$ and $E_4$ must be computed numerically. Henceforth $a$
will be the distance between the sphere and the plate and $R$ the
radius of the sphere.

Comparison with the parallel plate case as $a\to 0$ suggests the
error due to neglecting the fifth and higher reflections to be
$\sim 2\%$. Hence we have plotted our results as a band $2\%$ in
width in Fig.~2. Since the fractional contribution of higher
reflections decreases with $a$, we believe this is a conservative
estimate for larger $a$.

The proximity force approximation has been the standard tool for
estimating the effects of departure from planar geometry for
Casimir effects for many years\cite{MT}. In this approach one
views the sphere and the plate as a superposition of infinitesimal
parallel plates (i.e.\ the sphere is substituted by a stairlike
surface). The resulting expression is
\begin{equation}
    \cE_{\rm PFA}^{\rm plate} = -\frac{\pi^{3}\hbar cR}{1440 a^{2}}
    \frac{1}{1+a/R}.
    \label{PFApl}
\end{equation}
It is custom to factor out the most divergent term of the Casimir
\emph{force} in the limit $a/R\to 0$ as predicted by the PFA, in
this case $- \pi^{3}\hbar c R/720 a^{3}$, so to write in general
\begin{equation}
    \cF  = -f(\frac{a}{R})\frac{\pi^{3}\hbar c R}{720a^{3}},
    \label{ffunction}
\end{equation}
which is the definition of $f$. Modern experiments are approaching
accuracies where the deviations of $f(a/R)$ from unity are
important. PFA predicts that
\begin{eqnarray}
    f_{\rm PFA}(a/R) &=& 1-\frac{1}{2}\frac{a}{R} +
    {\it
    O}\left(\frac{a^{2}}{R^{2}}\right),
\label{PFAforcePlate}
\end{eqnarray}
while the optical approximation data predict
\begin{equation}
f_{\rm optical}(a/R)=1+0.05\ a/R+{\it
O}\left(\frac{a^{2}}{R^{2}}\right).
\end{equation}

Beyond the limit of small $a/R$, one must notice that the optical
approximation to the Casimir energy and the data of
Ref.~\cite{Gies03} both fall like $1/a^{2}$ at large $a/R$. In
fact both are roughly proportional to $1/a^{2}$ for all $a$. In
contrast the PFA estimates of the energy falls like $1/a^{3}$ at
large $a$ and departs from the Gies \emph{et al.}\cite{Gies03}
data at relatively small $a/R$.  For purposes of display we
therefore scale the estimates of the energy by the factor
$-1440a^{2}/\pi^{3}R\hbar c$.  The results are shown in Fig.~2.
\begin{figure}[ht]
\label{fig:esph}
\centerline{\epsfxsize=3.3in\epsfbox{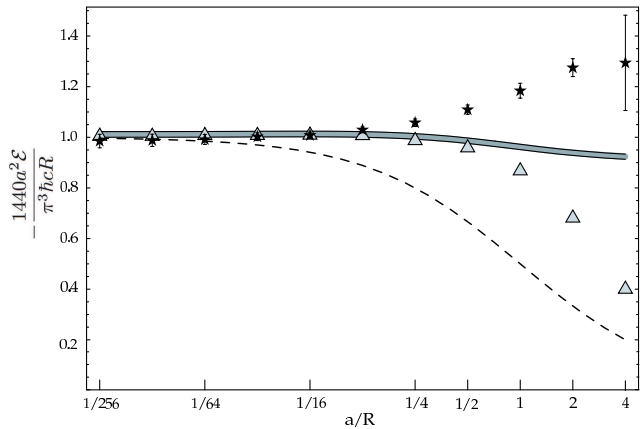}} \caption{Sphere
facing a plane case. Comparison between different methods. Results
for Ref.4 (stars with error bars), data from 6, superseded by this
work (triangles), optical approximation (thick grey line), PFA
(broken line).}
\end{figure}
The dominant contribution, always greater than 92\%, comes from
the second reflection.  The fourth reflection contributes about
6\% for $a/R\ll 1$ and less as $a/R$ increases. The contributions
of the first and third reflections are very small for all $a/R$. A
relevant result, confirmed by the analytical analysis on the
energy momentum tensor (within the optical
approximation\cite{opt2}) is that the asymptotic behavior of $\cE$
as $a/R\gg 1$ predicted by the optical approximation is $\propto
1/a^2$.  This is in contrast with the Casimir-Polder law which
predicts $E\propto 1/a^4$ at large $a$, the discrepancy is to be
attributed to settling in of diffraction effects.
%

\section{Conclusions}
\label{sec:concl}
We have proposed a new method for calculating approximately
Casimir energies between conductors in generic geometries.  We use
an approximation imported from studies of wave optics that we have
therefore named the ``optical approximation''. In this paper, we
have outlined the derivation and applied it to two examples: the
canonical example of parallel plates and the experimentally
relevant situation of a sphere facing a plane. Our results are in
agreement with the Proximity Force Approximation only to leading
order in the small distances expansion. The first order correction
is found to be different. This is of particular importance in the
example of the sphere and the plane because the first order
correction in $a/R$ ($a$ is the distance sphere-plate and $R$ is
the radius of the sphere) will soon be measured by new precision
experiments.

\vspace{0.2cm}

We would like to thank the organizers of the conference
`Continuous Advances in QCD 2004'. This work is supported in part
by the U.S.~Department of Energy (D.O.E.) under cooperative
research agreement~\#DF-FC02-94ER40818 and in part by the INFN-MIT
`Bruno Rossi' fellowship.


\end{document}